\documentstyle[12pt,epsfig]{article}

\makeatletter \renewcommand{\@biblabel}[1]{#1.} \makeatother

\begin{document}
\pagestyle{empty}
\renewcommand{\refname}{{\large \textbf{REFERENCES}}}


\begin{center}
{\large{ {\textbf{Defects, Tunneling, and EPR Spectra of 
Single-Molecule Magnets} }}}
\end{center}
\noindent{Kyungwha Park$^{1,2,3}$,
M.~A.\ Novotny$^4$, N.~S.\ Dalal$^3$, S.\ Hill$^5$, 
P.~A.\ Rikvold$^{1,6}$, S.\ Bhaduri$^7$, G.\ Christou$^7$, 
and D.~N.\ Hendrickson$^8$} \\
\noindent{$^1$School of Computational Science and
Information Technology, Florida State University, Tallahassee,
Florida 32306} \\
\noindent{$^2$Department of Chemistry and Biochemistry, 
Florida State University,
Tallahassee, Florida 32306} \\
\noindent{$^3$ Center for Computational Materials Science,
Code 6390, Naval Research Laboratory, Washington DC 20375} \\
\noindent{$^4$Department of Physics and Astronomy and the 
Engineering Research Center, Mississippi
State University, Mississippi State, Mississippi 39762} \\
\noindent{$^5$Department of Physics,
University of Florida, Gainesville, Florida 32611} \\
\noindent{$^6$Center for Materials Research and
Technology and Department of Physics,\\
Florida State University, Tallahassee, Florida 32306} \\
\noindent{$^7$Department of Chemistry, University of Florida, 
Gainesville, Florida 32611} \\
\noindent{$^8$Department of Chemistry and Biochemistry, University
of California at San Diego,\\ La Jolla, California 92093} \\

\noindent{\textbf{ABSTRACT}} \\

We examine theoretically electron paramagnetic resonance (EPR) lineshapes 
as functions of resonance frequency, energy level, and temperature
for single crystals of three different kinds of single-molecule 
nanomagnets (SMMs): 
Mn$_{12}$ acetate, Fe$_8$Br, and the $S=9/2$ Mn$_4$ compound.
We use a density-matrix equation and consider distributions in the uniaxial
(second-order) anisotropy parameter $D$ and the $g$ factor, caused by 
possible defects in the samples. Additionally, weak intermolecular exchange and 
electronic dipole interactions are included in a mean-field approximation. 
Our calculated linewidths are in good agreement with experiments.
We find that the distribution in $D$ is common to the three examined
single-molecule magnets. This could provide a basis for a proposed tunneling 
mechanism due to lattice defects or imperfections.
We also find that weak intermolecular exchange and dipolar interactions are 
mainly responsible for the temperature dependence of the lineshapes 
for all three SMMs, and that
the intermolecular exchange interaction is more significant for Mn$_4$ 
than for  
the other two SMMs. This finding is consistent with earlier experiments
and suggests the role of spin-spin relaxation processes in the mechanism
of magnetization tunneling. \\

\noindent{\textbf{INTRODUCTION}} \\

Single-molecule magnets (SMMs) have recently been the focus of 
much attention because of the possibility of 
macroscopic quantum tunneling of their magnetizations \cite{GUNT95,CHUD98} and 
possible applications in magnetic storage devices and quantum computers 
\cite{LEUE01}.
SMMs are composed of identical single-domain nanoscale molecules, 
comprised of a core of several transition-metal ions surrounded by 
many different species of atoms, and they have a large effective spin. 
The characteristics of SMMs are relatively weak exchange and dipolar
interactions between molecules, a large zero-field energy barrier 
against magnetization reversal, and magnetization 
steps in their hysteresis loops, 
which indicate quantum tunneling despite the large spin values
\cite{VILL94,BARR97,HILL98,PERE98,MIRE99}.

In this paper, we examine three different molecular magnets, which
are briefly described in the following.
The most extensively studied single-molecule magnet is 
[Mn$_{12}$O$_{12}$(CH$_3$COO)$_{16}$(H$_2$O$_4$)]$\cdot$2CH$_3$COOH$\cdot$4H$_2$O
(abbreviated hereafter as Mn$_{12}$), which was first synthesized by 
Lis \cite{LIS80}.
Each molecule in Mn$_{12}$ has an effective spin of $S=10$. 
It is a uniaxial system with  
a zero-field energy barrier against magnetization reversal 
of 65~K \cite{BARR97,HILL98,PERE98}. 
Another well-studied magnet is 
[Fe$_8$O$_2$(OH)$_{12}$(tacn)$_6$]Br$_8$$\cdot$9H$_2$O 
(abbreviated as Fe$_8$) \cite{WIEG84}. 
Each molecule in Fe$_8$ also has a spin of $S=10$
with a magnetization-reversal barrier of 30~K \cite{EDEL00,MACC01}. 
This is a biaxial system so exhibits oscillations in 
tunneling rates with transverse magnetic field \cite{GARG93}.
Recently synthesized is [Mn$_4$O$_3$(OSiMe)$_3$(O$_2$CEt)$_3$(dbm)$_3$] 
(abbreviated as Mn$_4$). Each molecule in Mn$_4$ consists 
of two kinds of Mn ions 
with mixed valence: three Mn$^{3+}$ ($S=2$) and one Mn$^{4+}$ ($S=3/2$) are
located in a distorted cubane structure and antiferromagnetically coupled 
to give a ground-state spin 
of $S=9/2$. The approximate magnetization-reversal 
energy barrier is 13~K in zero field \cite{EDWA02}.
All three of these SMMs show clear magnetization steps in hysteresis loops
below their blocking temperatures.

Recently, electron paramagnetic resonance (EPR) experiments
\cite{MACC01,EDWA02,PARK02-1,HILL02}
on single crystals of these SMMs showed interesting results 
for the lineshapes
as a function of EPR resonance frequency, energy levels involved in 
the EPR transitions,
and temperature. The temperatures employed are in the 2$-$50~K range, and
the frequencies are in the 40$-$190~GHz range.
We focus on the case when the magnetic field is applied along
the easy axis at fixed frequency. In this case, the measured linewidths 
are much broader than 
the homogeneous line broadening caused by the lifetimes of the energy levels, 
and the linewidths
increase as lower energy levels are involved with the EPR transitions. 
Additionally, some interesting features have been observed in the temperature
dependence of the linewidths and lineshapes of the EPR spectra
\cite{HILL98,MACC01,EDWA02,PARK02-1,HILL02}.
These temperature dependences were different for different molecular magnets.
In this paper, we summarize our theoretical understanding of the linewidths' 
dependence on frequency, energy level, and temperature for the
three molecular magnets, Mn$_{12}$, Fe$_8$, and Mn$_4$. We also provide 
quantitative comparison of our calculated results with experimental data.
The theoretical results on Mn$_4$ are preliminary and 
work on this system is still in progress. \\

\noindent{\textbf{MODEL}} \\

We start with a single-spin effective Hamiltonian considering
embedded symmetry in each molecule and calculate the resonance linewidths 
using a density-matrix equation \cite{BLUM96}. 
We find that an entirely single-spin picture 
cannot explain even qualitative trends in the experimental data. Thus, 
it is inevitable to include many-body effects.
As a simple start, we assume that the second-order uniaxial anisotropy
parameter, $D$, and the $g$ factor may not be the same for all molecules.
The microscopic origin of the distribution in $D$ has not
yet been fully understood \cite{CHUD01,CORN01}.  
In addition, each molecule may interact with the rest of the molecules
through exchange and/or dipolar interactions.
Details of the model and the technique are discussed
in Refs.~\cite{PARK02-1,PARK02-4,MCMI60,VANV48}.
This section is based in part on our earlier work \cite{PARK02-1,PARK02-4}.

Since each molecule in Mn$_{12}$ has approximate S$_4$ symmetry,
the dominant single-spin Hamiltonian reads, with applied magnetic field
along the easy magnetization axis ($z$ axis),
\begin{eqnarray}
{\cal H}_0 &=& -DS_z^2 -CS_z^4 - g\mu_B H S_z \;,
\end{eqnarray}
where $D=0.55$~K, $C=1.17 \times 10^{-3}$~K, and $g=1.94$ \cite{BARR97}. 
The SMM Fe$_8$ has approximate D$_2$ symmetry, so the Hamiltonian
reads
\begin{eqnarray}
{\cal H}_0 &=& -DS_z^2 -E(S_x^2 - S_y^2) - g\mu_B H S_z \;,
\end{eqnarray}
where $D=0.288$~K, $E=0.043$~K, and $g=2.0$ \cite{MACC01}.
In our calculations for Fe$_8$, we neglect the small $E$ term
because we are interested in the case with the applied field along
the easy axis.
The SMM Mn$_4$ has C$_3$ symmetry, so the Hamiltonian
reads
\begin{eqnarray}
{\cal H}_0 &=& -DS_z^2 -CS_z^4 - g\mu_B H S_z \;,
\end{eqnarray}
where $D=0.632$~K, $C=3.12 \times 10^{-3}$~K, and $g=2.0$ \cite{EDWA02}.

We now introduce an interaction $V(t)$ between the spin system and
an oscillating transverse magnetic field $H_x$ with angular
frequency $\omega\equiv 2\pi\nu$. The interaction of the
spin system with the environment is governed by
a time dependent density-matrix equation \cite{BLUM96,LEUE00}.
The power absorption between the energy levels $M_s$ and $M_s-1$
from the oscillating field, up to first order in $V_0$ and
near resonance (with a fixed value of $D$), is written as
\begin{eqnarray}
\frac{{\rm d}{\cal E}}{{\rm d}t}
&=&
\!\frac{({\cal E}_{M_s-1}-{\cal E}_{M_s})}{\hbar^2} 
|\langle M_s|V_0|M_s-1 \rangle|^2
\Delta(H) (\rho_{M_s}  -  \rho_{M_s-1}) \;, 
\label{pwa} \\
\Delta(H)&\equiv& \frac{\hbar^2 \gamma_{M_s-1,M_s}}
{ (g \mu_B)^2 ( H - H_{\rm res})^2 + ( \hbar \gamma_{M_s-1,M_s} )^2 } \;, 
\label{eq:loren} \\
H_{\rm res} &\equiv& \frac{\hbar \omega - D(2 M_s - 1)
- C(4 M_s^3 - 6 M_s^2 + 4 M_s -1 )}{g \mu_B} \;,
\label{eq:res_field}
\end{eqnarray}
where $V_0$ is the strength of the interaction $V(t)$,
${\cal E}_{M_s}$ is the energy of the level $M_s$,
$\rho_{M_s}$ is the population of the level $M_s$,
$\Delta(H)$ is a Lorentzian lineshape function,
and $H_{\rm res}$ is the resonance field for Mn$_{12}$ and Mn$_4$.
(For Fe$_8$ there is no $C$ term.)
The ratio $\hbar \gamma_{M_s-1,M_s}/g \mu_B$ gives a linewidth
due to the finite lifetime of any excited state.
The linewidths determined by $\gamma_{M_s-1,M_s}$ are on the
order of several to several tens of gauss at the measured
temperature (10 K for Fe$_8$ and Mn$_4$ and 25 K for Mn$_{12}$),
and they decrease with increasing $M_s$. (In our
convention, $M_s=+10$ is the ground state with $H>0$.)
However, the measured linewidths {\it increase} with increasing $M_s$, and
the order of magnitude of the widths is a couple of hundred
to a thousand gauss. Additionally, the dramatic temperature
dependence of the linewidths cannot be explained
by homogeneous broadening alone.

To resolve this discrepancy, we include distributions in $D$ and $g$,
(called $D$-strain and $g$-strain effects, respectively)
and intermolecular exchange and dipolar interactions.  
With Gaussian distributions in $D$ and $g$, centered at the experimentally 
determined values and 
with small standard deviations $\sigma_D$ and $\sigma_g$,
we can average the power absorption (equation \ref{pwa}) over the
distributions. To calculate the effect on the linewidths of
the intermolecular exchange and dipolar interactions with fixed $D$ and $g$, 
we use a multispin Hamiltonian \cite{PARK02-4,MCMI60,VANV48}:
\begin{eqnarray}
{\cal H}^{\mathrm tot}&=&\sum_{i}[\: {\cal H}_{0i} + V_{i}(t) \:] \:
+ {\cal H}^{(1)} \: , \: \: \: {\cal H}^{(1)} \equiv
{\cal H}^{\mathrm dipole} + {\cal H}^{\mathrm exch} \:, \\
{\cal H}^{\mathrm dipole}&=&\frac{1}{2} \sum_{jk}\,^{\prime} A_{jk}
( \vec{S}_j \cdot \vec{S}_{k} - 3  S_{j}^z S_{k}^z ) \:, \\
\label{eq:dp}
A_{jk} &\equiv& \left( \frac{\mu_0}{4 \pi} \right)
\frac{(g \mu_B)^2}{2 r_{jk}^3} (3 \zeta_{jk}^2 - 1) \:, \\
{\cal H}^{\mathrm exch}&=& \: \frac{1}{2} \sum_{jk}\,^{\prime}
J_{jk} \: \vec{S}_j \cdot \vec{S}_{k} \:,
\label{eq:ex}
\end{eqnarray}
where ${\cal H}_{0i}$ is the single-spin Hamiltonian for the $i$th
molecule, the sum $\sum_{i}$ runs over all molecules,
and $V_i(t)$ is the interaction
between the $i$th molecule and the oscillating transverse field. Here
${\cal H}^{\mathrm dipole}$ is the dipolar interaction between the molecules,
and $\zeta_{jk}$ are the direction cosines of the vector
between molecules $j$ and $k$ ($\vec{r}_{jk}$) relative to the
easy axis ($z$ axis). The sum $\sum_{jk}^{\prime}$ runs over
all molecules, so that any two indices are not the same.
${\cal H}^{\mathrm exch}$ is the isotropic
exchange interaction between the spins of nearest-neighbor molecules,
where the exchange coupling constant $J_{ij}$ is $J$
if the $i$th and $j$th spins are nearest neighbors and zero otherwise.
Assuming that $\sum_{i} V_i(t)$ is much smaller than ${\cal H}^{(1)}$, 
which again is much smaller than $\sum_i {\cal H}_{0i}$, we 
neglect $\sum_{i} V_i(t)$ and treat ${\cal H}^{(1)}$ perturbatively.
In a mean-field approximation, the sums of $A_{jk}$ and $J_{jk}$
(equations \ref{eq:dp} and \ref{eq:ex}) can be separated from the
spin operators.

Using the multispin Hamiltonian, we construct a probability
density function for the resonance field. 
To compare with the measured linewidths, 
we calculate the second 
central moment of the resonance field,
$\langle (H - \langle H \rangle)^2 \rangle$,
to zero order in ${\cal H}^{(1)}/k_B T$. To explain the experimental data,
we need to vary two parameters for the $D$-strain and $g$-strain effects,
and to vary three parameters \cite{PARK02-4} for the contribution of
the spin-spin interactions: $\sum_{ij}^{\prime} J_{ij}^2$,
$\Gamma\equiv \sum_{ij}^{\prime} A_{ij}^2$, and 
$\Lambda\equiv \sum_{ij}^{\prime} J_{ij} A_{ij}$. \\

\noindent{\textbf{RESULTS AND DISCUSSION}} \\

We find that the uniaxial anisotropy parameter $D$ is randomly distributed
in all three of the molecular magnets. This may be due to possible defects
or imperfections in the the samples. 
For Mn$_{12}$ and Mn$_4$, the distribution in 
the $g$ factor is also important, although this effect is not significant
for Fe$_8$. The spreads in $D$ and $g$ vary from sample to sample by 30$-$40\%.
The effect of the distribution in $D$ on inhomogeneous broadening was 
observed in terahertz spectroscopy for Mn$_{12}$ \cite{PARKS01}.
The temperature dependence of the linewidths is mainly caused by the spin-spin
interactions. The dipolar field was measured
in a millimeter-wave study of Fe$_8$ \cite{MUKH01}.
For Mn$_4$, the intermolecular exchange interaction is 
stronger than the dipolar interaction, which is consistent with experiments
on Mn$_4$ dimers \cite{WERN02-1} and monomers \cite{WERN02-2,WERN02-3}.
For clarity, we discuss each SMM separately. \\

\noindent{\textbf {\underline{Fe$_8$}}} \\

We find that for Fe$_8$ the distribution in $D$ and the intermolecular exchange
and dipolar interactions contribute to the linewidths.
The distribution in $D$ makes the linewidth increase with 
increasing energy level $M_s$, which can be seen from 
the resonance-field expression (equation \ref{eq:res_field}). 
The intermolecular exchange and dipolar interactions make the
linewidths increase with increasing $M_s$ and with decreasing
the resonance frequency $\nu$. The reason is that the 
resonance field decreases with increasing $M_s$ and 
with decreasing $\nu$, which can be understood by equation
\ref{eq:res_field}. Combining the effect of the distribution in $D$ with
the exchange and dipolar interactions explains well the trend
and the magnitude of the measured 
linewidths as functions of frequency and energy level \cite{PARK02-1}.

\begin{figure}[bt]
\begin{center}
\resizebox{6.cm}{6.cm}{\includegraphics{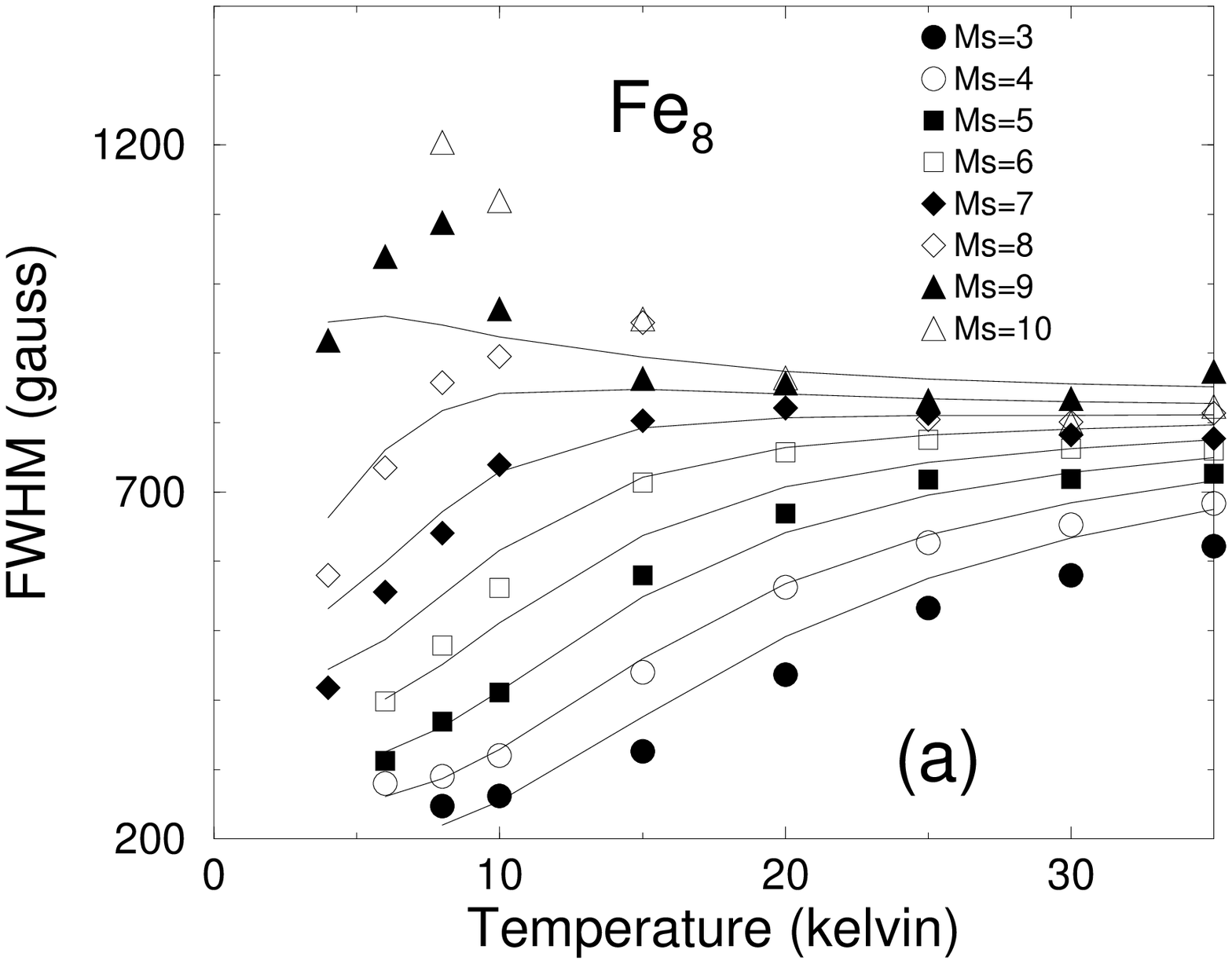}} 
\resizebox{6.cm}{6.cm}{\includegraphics{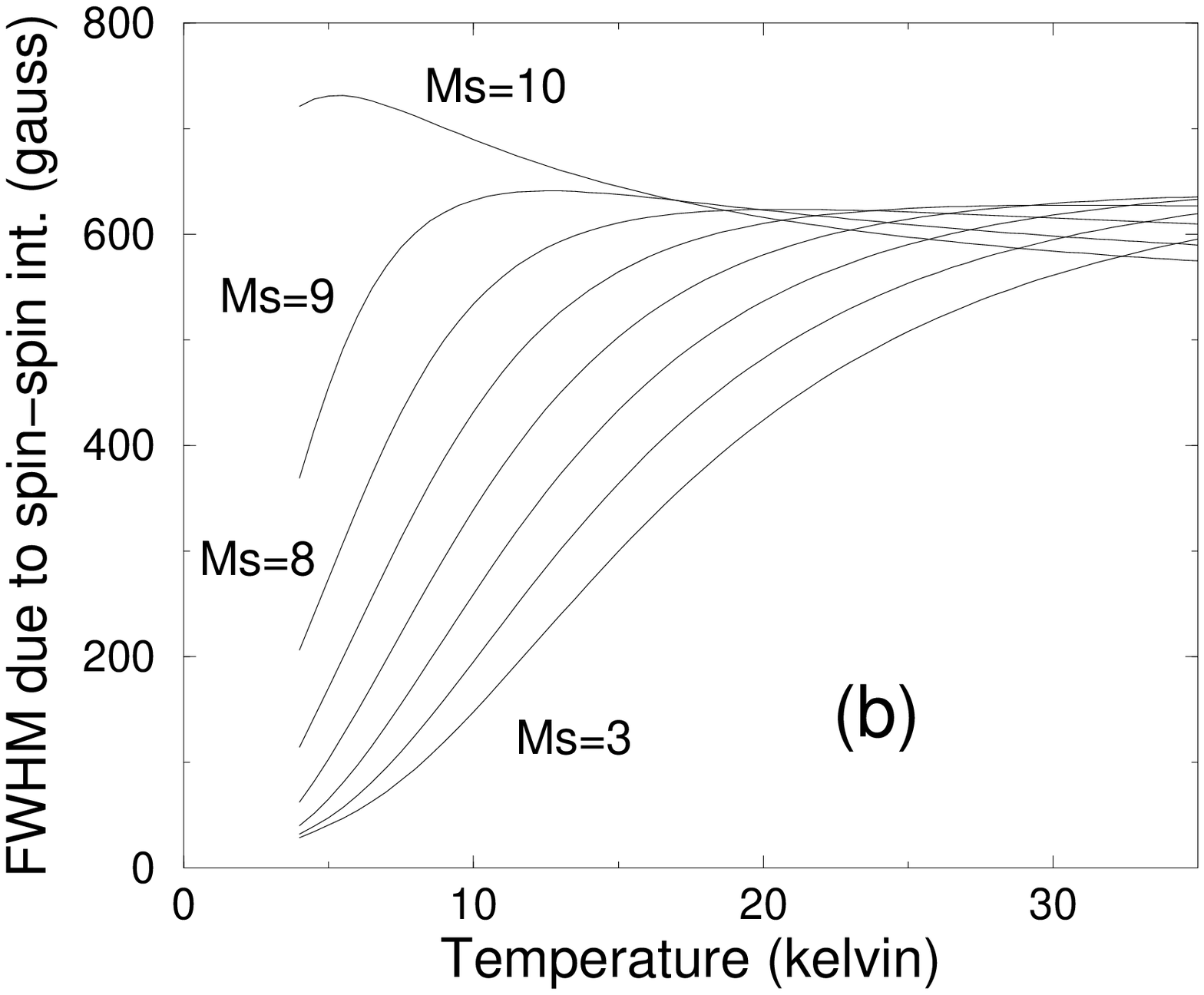}} 
\caption{(a) Calculated (curves) and measured (symbols) FWHM
vs temperature at $\nu=116.9$~GHz for Fe$_8$. Here
the standard deviation of $D$, $\sigma_D\sim 0.0064D$, the exchange
constant $J=-7$~G, $\Gamma=\sum_{ij}^{\prime} A_{ij}^2/N =86$~G$^2$,
and $\Lambda=\sum_{ij}^{\prime} J_{ij} A_{ij}/N =-156$~G$^2$.
The solid curves, from bottom to top, correspond to $M_s=3,4,...,9,10$.
(b) Calculated FWHM caused by the exchange and dipolar interactions only,
shown vs temperature at $\nu=116$~GHz for Fe$_8$. Here
$J=-7$~G, $\Gamma=86$~G$^2$, and $\Lambda=-156$~G$^2$. 
Cited from Ref.~\protect{\cite{PARK02-4}}.}
\label{fig:Fe8WTdep}
\end{center}
\end{figure}

Next we discuss the temperature dependence of the linewdiths.
The distribution in $D$ alone cannot explain the interesting 
measured temperature dependence of the linewidths shown
as symbols in figure \ref{fig:Fe8WTdep}~(a). The reason is that
the linewidths caused by the distribution in $D$ are only
slightly temperature dependent due to the temperature dependence
of the natural linewidths $\hbar \gamma_{M_s-1,M_s}/g \mu_B$ 
in equation \ref{eq:loren}. So this weak temperature dependence
is monotonic and noticeable only for the small $M_s$ (at most
about 100~G at 30~K for $M_s=3$) \cite{PARK02-4}.
However, the linewidths caused solely by the exchange and dipolar interactions 
at fixed $D$ and $g$ vary significantly with temperature, as shown 
in figure \ref{fig:Fe8WTdep}~(b). For the ground state $M_s=10$,
the linewidths decrease with increasing temperature in the
whole temperature range. For $M_s=9$, 8, and 7, the widths first increase
sharply with temperature at low temperatures, and then
decrease slowly with temperature at high temperatures. For $M_s=6$, 5, 4,
and 3, the widths increase with increasing temperature in the whole
range. This trend was also seen in the experimental
linewidths, confirming that the exchange and dipolar interactions 
are crucial to understanding the temperature dependence of the linewidths.
The calculated linewidths including the distribution in $D$ and the
spin-spin interactions are shown as curves in figure \ref{fig:Fe8WTdep} (a).
Here we use $\sigma_D \sim 0.0064\;D$, $J=-7$~G, $\Gamma=86$~G$^2$,
and $\Lambda=-156$~G$^2$. The calculated linewidths agree with
the measured linewidths except in the low-temperature range
for large $M_s$ transitions ($M_s=10$, 9, 8). At present,
we do not fully understand this discrepancy.
We note that the quality of the fit that includes
a small ferromagnetic exchange interaction is comparable
with that without exchange interaction, because the 
exchange coupling constant is small compared with
the linewidths\cite{AABB}. \\

\noindent{\textbf {\underline{Mn$_{12}$}}} \\

\begin{figure}[bt]
\begin{center}
\resizebox{6.cm}{6.cm}{\includegraphics{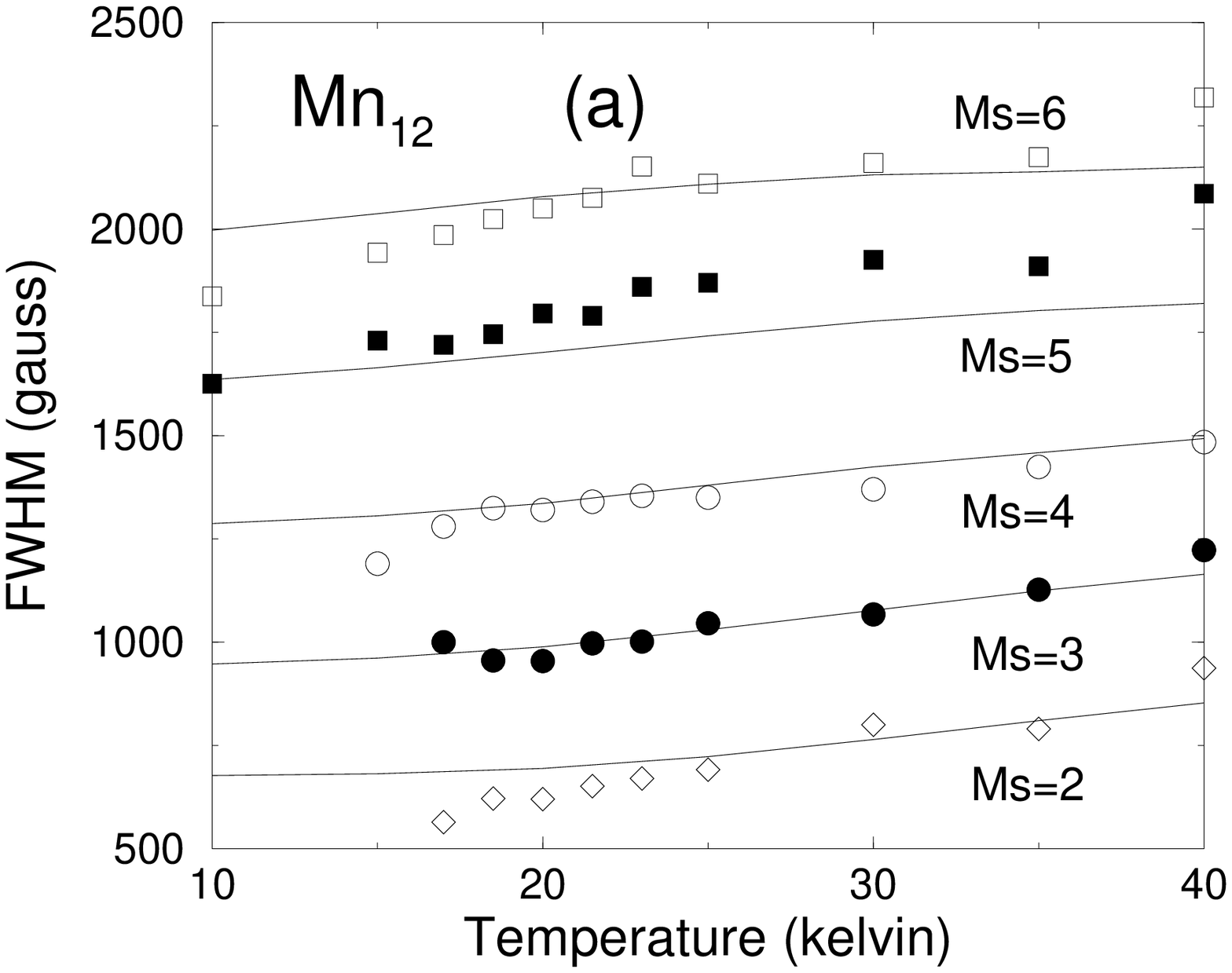}} 
\resizebox{6.cm}{6.cm}{\includegraphics{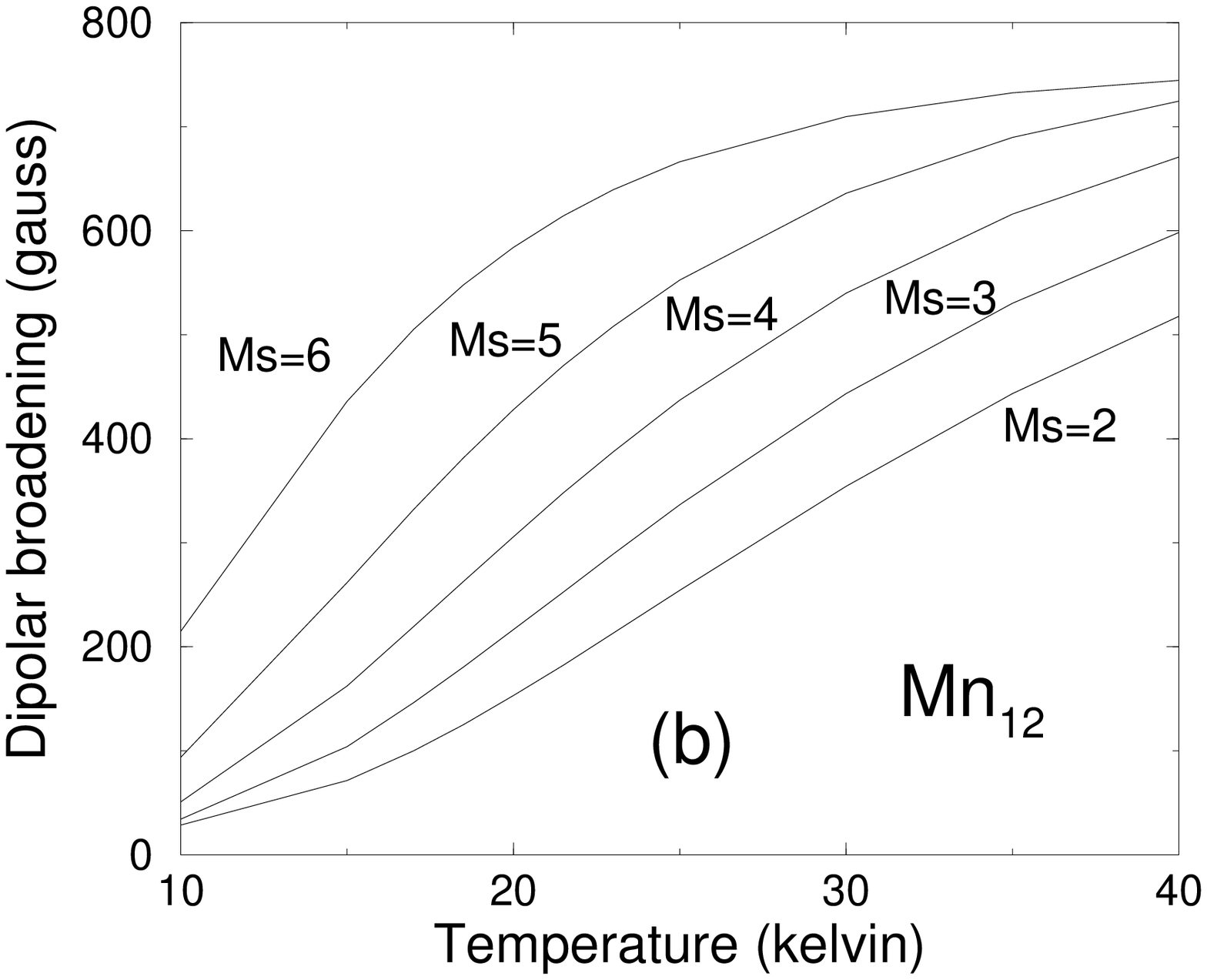}}
\caption{(a) Calculated (curves) and measured (symbols) FWHM vs temperature
at $\nu=189.123$~GHz for Mn$_{12}$.
Here the $D$ strain ($\sigma_D\sim 0.018D$), $g$ strain ($\sigma_g\sim 0.002g$),
and the dipolar interactions ($\Gamma = 203$~G$^2$) are included in
the calculated linewidths.
(b) Calculated FWHM caused by the dipolar interactions only, shown
vs temperature, at $\nu=189.123$~GHz with $\Gamma=203$~G$^2$ for Mn$_{12}$.
The examined temperature range for Mn$_{12}$ is from 10~K to 40~K.
Cited from Ref.\protect{\cite{PARK02-4}}. }
\label{fig:Mn12Tdep}
\end{center}
\end{figure}

We find that the inhomogeneous line broadening for Mn$_{12}$
is due to the distributions in $D$ and $g$, and to the
dipolar interactions. The distribution in $D$ contributes
to the variation of the linewidth as a function of $M_s$ and $\nu$ in the
same way as for Fe$_8$. The distribution in $g$ makes
the linewidths {\it decrease} with increasing $M_s$
and decreasing $\nu$, which is opposite to the effect of 
the distribution in $D$.
The reason is that the resonance field
decreases with increasing $M_s$ and decreasing $\nu$.
The temperature dependence of the linewidths caused
by only the distributions in $D$ and $g$ is very weak,
so that it cannot explain the measured linewidths
shown as symbols in figure \ref{fig:Mn12Tdep} (a),
which is similar to those for Fe$_8$.
The contribution of the dipolar interactions to the linewidths
is shown as a function of temperature in figure \ref{fig:Mn12Tdep} (b).
Unlike for Fe$_8$, the $M_s$ dependence of the dipolar broadening does
not decrease with increasing temperature (the curves are almost
parallel as the temperature increases).
Combining the three effects (distributions in $D$ and $g$,
and dipolar interactions), we find that the calculated
linewidths agree well with the experimental data with
$\sigma_D\sim 0.018D$, $\sigma_g\sim 0.002g$, and $\Gamma=203$~G$^2$,
as shown in figure \ref{fig:Mn12Tdep}~(a). 
Compared to the linewidths for Fe$_8$, the relatively weak
temperature dependence for Mn$_{12}$ indicates that 
for Mn$_{12}$ the dipolar broadening is overshadowed by the
effect of the distribution in $D$, which is three times
as wide as for Fe$_8$. 
With the same parameter values, we also check that the $M_s$
and frequency dependence of the measured linewidths can be explained. \\

\noindent{\textbf {\underline{Mn$_4$}}} \\

We find that for Mn$_4$ all of our considerations of the many-body effects
- the distributions in $D$ and $g$, and the intermolecular
exchange and dipolar interactions - substantially affect the
inhomogeneous broadening. One minor difference of this SMM from the
other two SMMs is that the effect of the distribution in $D$ 
disappears in the transition $M_s=1/2 \rightarrow -1/2$ because
the resonance field (equation \ref{eq:res_field}) 
does not depend on $D$ at this transition.
To examine the temperature dependence of the linewidths
caused solely by $D$-strain and $g$-strain, we used
the same spin-phonon coupling parameters as for Mn$_{12}$\cite{LEUE00}
to obtain the order of magnitude of the lifetimes of the energy levels.
As expected from the analysis of the two other SMMs, 
we also find that for Mn$_4$ the distributions in $D$ and $g$ do not
significantly change the dependence of the linewidths on temperature.

\begin{figure}
\begin{center}
\resizebox{6.cm}{6.cm}{\includegraphics{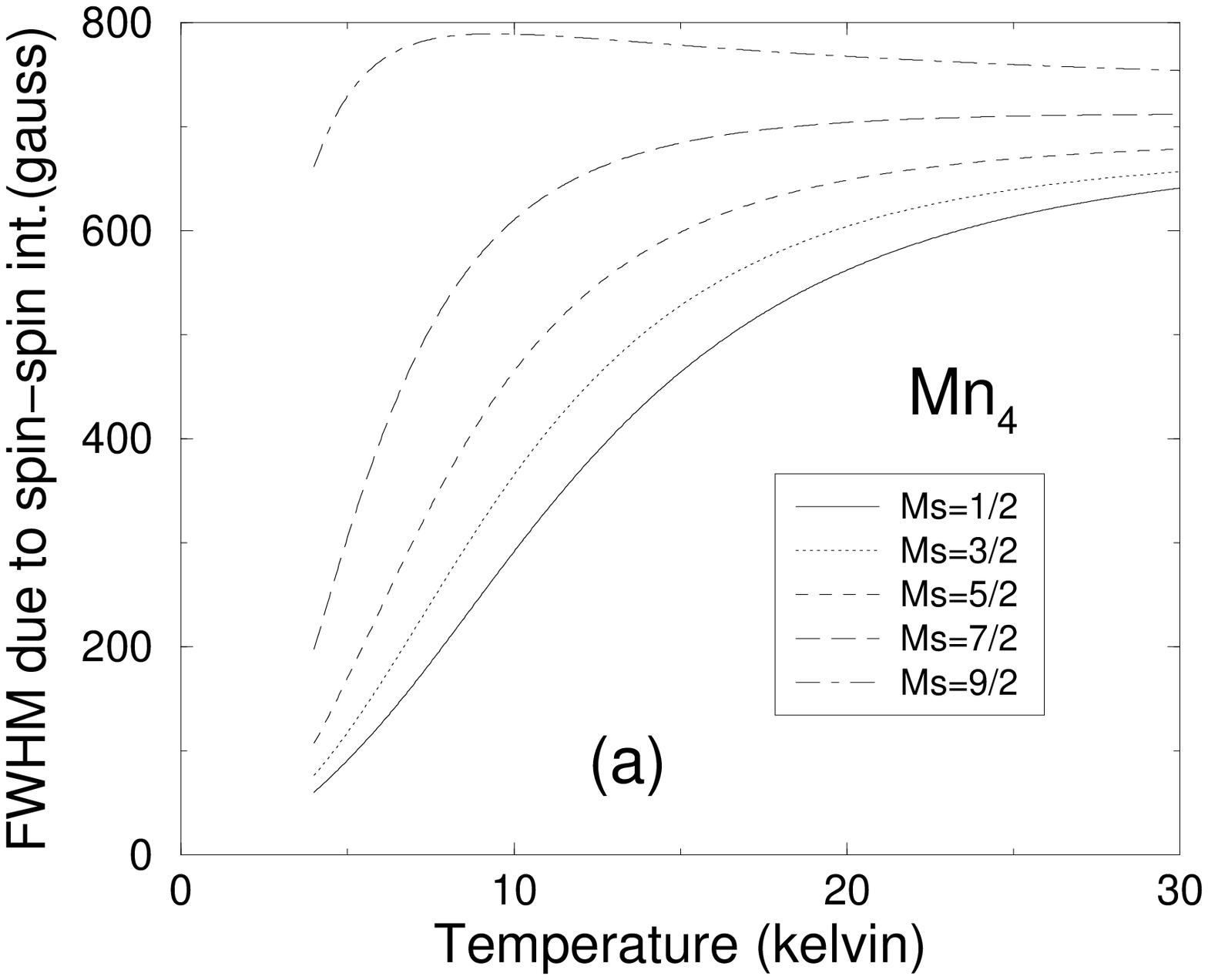}} 
\resizebox{6.cm}{6.cm}{\includegraphics{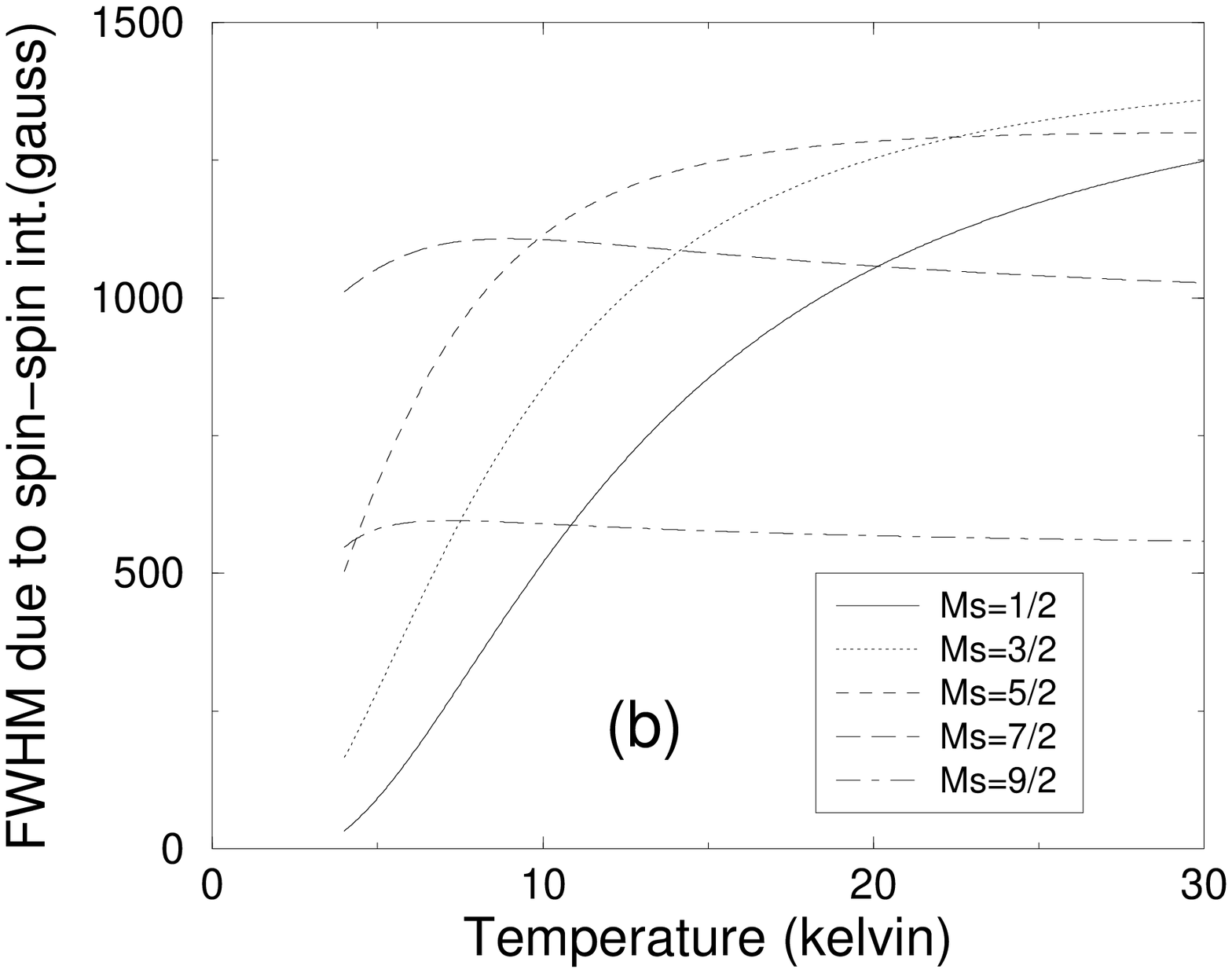}}  
\caption{(a) Calculated linewidths due to the exchange and dipolar 
interactions only
vs temperature at $\nu=138$~GHz for Mn$_4$. Here we use the exchange 
coupling constant $J=-40$~G, $\Lambda=\sum_{ij}^{\prime} J_{ij} A_{ij}/N 
= -1854.8$~G$^2$, and $\Gamma=\sum_{ij}^{\prime} A_{ij}^2/N = 344$~G$^2$.
(b) The same linewidths as in (a) when the sign of $J$ is flipped and
the other parameter values are kept the same.}
\label{fig:Mn4spin}
\end{center}
\end{figure}

Figure \ref{fig:Mn4spin}~(a) 
shows the calculated linewidths caused by the exchange
and dipolar interactions only at fixed $D$ and $g$ versus temperature at
$\nu=138$~GHz. Here we use the exchange coupling constant $J=-40$~G,
two nearest neighbors along the easy axis (because the nearest-neighbor 
distance along this direction is half of the nearest-neighbor 
distance perpendicular to this direction), $\Lambda=
\sum_{ij}^{\prime} J_{ij} A_{ij}/N = -1854.8$~G$^2$, 
and $\Gamma=\sum_{ij}^{\prime} A_{ij}^2/N = 344$~G$^2$. 
The negative sign in $J$ means ferromagnetically
coupled spins. If we assume the approximate lattice geometry
as a tetragonal lattice and take the easy axis as the crystal $c$ axis, 
then we know that the sign of $\Lambda$ should be the same as that of $J$.
We explore some other possibilities of combining the exchange with the dipolar
interaction. For example, figure \ref{fig:Mn4spin} (b) shows the linewidths 
due to positive $J$ (antiferromagnetically coupled) and the dipolar interaction. 
With positive $J$, the linewidth for $M_s=9/2$ is greatly reduced and the 
linewidth for $M_s=1/2$ is appreciably enhanced. This feature is quite
different from the broadening with negative $J$.

\begin{figure}
\begin{center}
\resizebox{7.cm}{7.cm}{\includegraphics{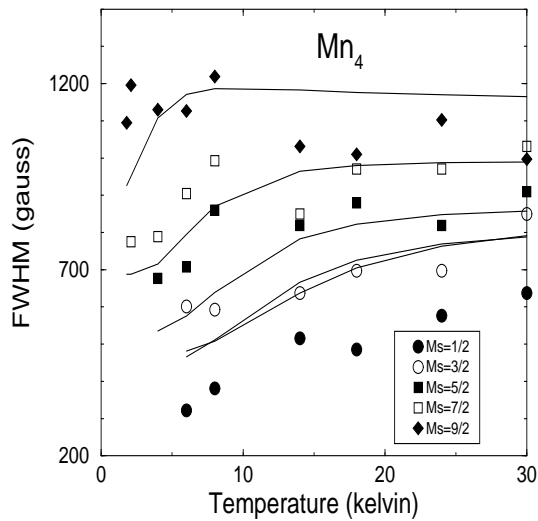}}  
\caption{Calculated (curves) and measured (symbols) FWHM vs temperature
at $\nu=138$~GHz for Mn$_4$. Here we use $\sigma_D=0.01D$, $\sigma_g=0.004g$,
$J=-40$~G, $\Lambda=\sum_{ij}^{\prime} J_{ij} A_{ij}/N = -1854.8$~G$^2$, 
and $\Gamma=\sum_{ij}^{\prime} A_{ij}^2/N = 344$~G$^2$.
The solid curves, from bottom to top, correspond to $M_s=1/2,3/2,...,9/2$.}
\label{fig:Mn4Tdep}
\end{center}
\end{figure}

Figure \ref{fig:Mn4Tdep} shows the calculated linewidths 
versus temperature at $\nu=138$~GHz.
Here we use $\sigma_D=0.01D$, $\sigma_g=0.004g$, $J=-40$~G,
$\Lambda=-1854.8$~G$^2$, and $\Gamma=344$~G$^2$. The calculated
linewidths agree reasonably with the experimental data except for the
transition $M_s=1/2 \rightarrow -1/2$. Notice that the magnitudes of the
values of $J$ and $\Lambda$ are greatly enhanced compared to
those for Fe$_8$. This suggests that Mn$_4$ has a stronger exchange
interaction between molecules than Fe$_8$. Our value
for $J$ is different by a factor of 2 
from the experimentally extracted value, $J=-74$~G \cite{WERN02-2}. 
Therefore, our finding may support the proposed
mechanism of spin-spin cross relaxation in Mn$_4$ \cite{WERN02-2}.

\begin{figure}
\begin{center}
\resizebox{7.cm}{7.cm}{\includegraphics{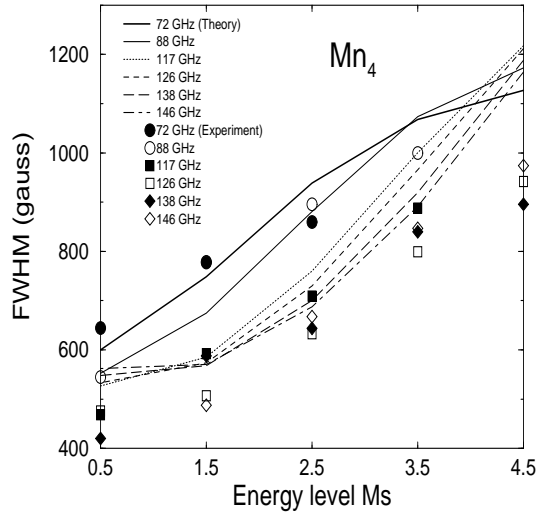}}  
\caption{Calculated (curves) and measured (symbols) FWHM vs energy level $M_s$
at $T=10$~K for $\nu=72$, 88, 117, 126, 138, and 146~GHz for Mn$_4$.
Here the values of $\sigma_D$, $\sigma_g$, $J$, $\Gamma$, and $\Lambda$
are the same as those in figure \ref{fig:Mn4Tdep}. }
\label{fig:Mn4Fdep}
\end{center}
\end{figure}

With the same parameter values, we can 
explain the frequency dependence of the linewidths 
at $T=10$~K shown in figure \ref{fig:Mn4Fdep}. Although there are
substantial distributions in $D$ and $g$, the widths are somewhat dependent
on the frequency, which indicates that the spin-spin interactions are
not completely overshadowed by the $D$-strain and $g$-strain effects
as in Mn$_{12}$. But definitely the frequency dependence is 
stronger than for Mn$_{12}$ but weaker than for Fe$_8$. \\

\noindent{\textbf{CONCLUSIONS}} \\

We investigated the EPR lineshapes as functions of resonance
frequency $\nu$, energy level $M_s$, and temperature $T$ for
the three different molecular magnets, Mn$_{12}$, Fe$_8$, and Mn$_4$,
when the magnetic field is applied along the easy axis and the frequency 
is kept fixed. In our calculations, intermolecular exchange
and dipolar interactions were included, as well as distributions in the uniaxial
anisotropy parameter $D$ and the $g$ factor. We find that
the distribution in $D$ is present in all three SMMs but that
the effect is strongest in Mn$_{12}$, and that
the distribution in $g$ contributes to the linewidths for Mn$_{12}$ and Mn$_4$.
The spin-spin interactions are responsible for the interesting trend 
of the temperature dependent lineshapes for all three magnets.
For Mn$_4$, the exchange interaction is stronger than the dipolar interaction,
which supports the conclusions derived from the earlier measurements\cite{WERN02-2}.
The linewidths for the three magnets reveal different 
$M_s$, $\nu$, and $T$ dependences
because of different contributions from the effects we are considering.
Observation of 
EPR lineshapes could indicate which effects are dominant in a particular 
SMM, so that we may identify which mechanism leads to quantum tunneling. \\

\noindent{\textbf{ACKNOWLEDGMENTS}} \\

This work was funded by NSF Grant Nos.~DMR-9871455, DMR-0120310,
DMR-0103290, and DMR-0196430, Research Corporation (S.H.),
and by Florida State University through the School of
Computational Science and Information Technology and
the Center for Materials Research and Technology. \\


\end{document}